\documentclass{sig-alternate-per}

\usepackage{amsmath}

\usepackage{cite}

\usepackage{url}
\usepackage{lipsum}
\usepackage{filecontents}
\usepackage{booktabs}
\usepackage{array}
 \usepackage{pgfplots}
\usepackage{pgfplotstable}

\usepackage{color}

\newcommand{\be}{\begin{equation}}
\newcommand{\ee}{\end{equation}}

\definecolor{co}{rgb}{0.8,0,0.8}
\definecolor{gr}{gray}{0.5}

\catcode`\@=11

\catcode`\@=12

\usepackage{comment}

\begin{document}

 \conferenceinfo{Symposium on Cryptocurrency Analysis (SOCCA) 2020}{~~~Milan, Italy}

\title{Crypto-Hotwire: Illegal Blockchain Mining at Zero Cost Using Public Infrastructures}


%
%
%
%
%
%
%
%

\author{Felipe Ribas Coutinho \\
\affaddr{UFRJ} \\
\affaddr{Rio de Janeiro, Brazil} \\
\and
Victor Pires \\
\affaddr{UFRJ} \\
\affaddr{Rio de Janeiro, Brazil} \\ 
\and
Claudio Miceli \\
\affaddr{UFRJ} \\
\affaddr{Rio de Janeiro, Brazil} \\
\and
Daniel S. Menasché \\
\affaddr{UFRJ} \\
\affaddr{Rio de Janeiro, Brazil} \\
}


\maketitle

\begin{abstract}
Blockchains and cryptocurrencies disrupted the  conversion of  energy into a medium of exchange. Numerous applications for blockchains and cryptocurrencies are now envisioned for purposes ranging from inventory control to banking applications. Naturally, in order to mine in an economically viable way, regions where energy is plentiful and cheap, e.g., close to hydroelectric plants, are sought. The possibility of converting energy into cash, however, also opens up opportunities for a new kind of cyber attack aimed at illegally mining cryptocurrencies by stealing energy. In this work, we indicate, using data from January and February of 2018 from our university, that such a threat is real, and present a projection of the gains derived from these attacks.

\end{abstract}






\section{Introduction}
With the popularization of virtual currencies (also known as cryptocurrencies), a new class of cyberattacks has emerged:  cryptojacking.  It consists of infecting and controlling the use of CPU and memory of computers for purposes of  mining cryptocurrencies.    According to~\cite{segura}, in 2018 such attacks were already more prevalent than classic attacks, such as DDoS, in certain regions of the US~\cite{govtech, free, college, vectra, coindesk}.

Mining attacks previously reported in the literature~\cite{hong2018you, kharraz2019outguard,ruth2018digging, segura, Zimba:18, pastor2020detection}  typically account for hacked  computers  being remotely exploited  and  locally monitored. We are unaware of  reports on the use of public computing infrastructures in which machines can be locally exploited  and remotely monitored.  This occurs, for instance, in universities and coffee shops, where a vast number of users has physical  access to machines,  posing additional challenges to their  remote   monitoring, e.g., by the university information technology (IT) team~\cite{govtech, free, college, vectra, coindesk}.

We refer to this new type of attack   as   ``crypto-hotwire'', in a reference to 
criminals that bypass meters and  ``hotwire'' the power supply  to  cut bills~\cite{BBC}.  Note that physical ``hotwire'' is typical in countries such as Brazil,  evidencing problems of public order which lend  energy theft as one of the major drivers for smart grids.  The theft of energy may occur for a number of  undue purposes. In the case of crypto-hotwire, the theft of energy occurs for  the mining of virtual currencies, which serves as a general medium of exchange.

Public infrastructures, in particular, are major targets of crypto-hotwire, since they have a large number of machines, significant Internet bandwidth and abundant energy. 
The large number of machines makes it easier for users to physically access the resources and  turn miners on and off.  
In contrast, to make use of home users' machines, e.g.,  through a botnet, such household machines need to have vulnerabilities that allow the attacker to invade the system remotely.  The attack complexity is   reduced when attackers can have physical access to a machine to install a mining module.

Once mining begins on a public infrastructure machine,  it leaves almost no trace on the network, rendering its detection very challenging. This is because simply monitoring the network   may not be enough to identify attacks, since they have a strongly local aspect and do not leverage any explicit  software vulnerabilities.  As a case study, in this work we study the action of attackers who seek to invade our university's machines to execute mining processes. This study aims to $(a)$ assess the prevalence of mining attacks at the university and  $(b)$  analyze the data generated by incidents, such as traffic and energy consumption, in order to generate a behavior pattern for detecting this type of attack.


In summary, the main contributions of this article are:

\textbf{We found that illegal cryptocurrency mining is prevalent: } Using data collected at the Federal University of Rio de Janeiro (UFRJ), we found that illegal cryptocurrency mining   occurs in practice. In particular, we were able to identify dozens of attempts of mining activities targeting  the Monero cryptocurrency. Unlike classic attacks that leave a trace on the bandwidth of the network, cryptocurrency mining can be very difficult to detect because it involves virtually only local activity at the affected machine.

\textbf{We quantify the costs and gains from illegal mining:} Through controlled lab  experiments, we quantify the costs and gains arising from the mining. In particular, we find that costs can escalate more quickly that earnings. Such a finding may motivate a wide increase in illegal use of resources. 


The remainder of this paper is organized as follows. After providing basic background, we present  our research methodology in Section~\ref{sec:method}. Sections~\ref{sec:finding1} and~\ref{sec:finding2} present our main findings, on the prevalence and gains from mining, and Section~\ref{sec:conclusion} concludes.


\section{Mining and mining attacks}
 
 The process of generating cryptocurrencies on a network is called mining.  This process consists of solving a mathematical problem (puzzle) and being rewarded for that solution.
   Mining demands high energy and power consumption for solving the puzzles. In order for mining to become a profitable process for the miner, it is necessary to reduce these costs. In an extreme situation, the miner can reduce  personal costs to zero, making use of energy from public infrastructure. If this is the case, the costs are incurred by those who finance such infrastructures.

    The mining process can be carried out on a single computer or using making such a group. In “solo” mining, the reward generated by the mining depends exclusively on the processing power of the computer. In pool mining, many computers mine together and the accumulated processing power  increases the chances of block resolution. When the block is resolved, the response compensation is proportionally divided according to the power used by each one. Currently, there are tools that allow any user to easily connect a machine to a mining pool. A user who has physical access to public infrastructure machine can connect it to a mining pool and individually profit from it.

\textbf{Traditional mining attacks. }
Next,  we present a brief introduction to mining attacks, indicating their classic types and what distinguishes them when the focus is on public infrastructure.  A mining or cryptojacking attack consists of the infection of one or more hosts, inserting in them malware that runs cryptocurrency mining software. 
It is worth mentioning that in this work we consider users who intentionally mine virtual coins on local physical machines. Alternatively, in the literature the focus has been on remote users performing such unwanted mining. This occurs in two ways: via Javascript or worms. In the first case, the mining code is executed in the browser and the loading of this code is done along with the loading of the desired page by the user, in a covert way. In the case of  worms, the executed code  is browser independent, potentially being downloaded individually.  

In the case of Javascript, monitoring CPU usage and memory of the browser, one can identify potential problems, while in the case of  worms, it is necessary to monitor the CPU and memory usage of the machine as a whole, as it is not known in advance what process is doing  malicious executions.   Such solutions are effective for home users, but they do not easily scale. Anomalies in the use of CPU and memory at the scale of universities, for instance, should be much harder to track than anomalies in a local machine. 



\textbf{New attack vectors and access vectors. } 
Although the topic of cryptojacking is in vogue~\cite{segura, hong2018you}, we are unaware of previous work that has analyzed the risks associated with this type of attack in public infrastructures, in which users have physical access to devices.  Anecdotal evidence suggests that those attacks are prevalent~\cite{coindesk, vectra, college}.


In related works, the authors consider that the browser does not intentionally access mining scripts (i.e., it is assumed a remote access vector) and that the honest user is interested in monitoring (locally) for suspicious activity. In this work, on the other hand, we assume that  the user intentionally executes mining scripts (local access vector, in physical facilities) and that monitoring needs to be remote, e.g., through deep packet inspection.


The problem considered in this work is particularly complex in developing countries, such as Brazil. In these countries, we have a significant portion of the population accessing the electrical network by means of physical hotwires. Such physical hotwires potentially gives rise to energy theft. 
The crypto-hotwire described in this work can be appreciated as a sophisticated version of the physical energy theft which are prevalent in slums. 
Although substantial resources  have been invested to detect such physical energy theft, they are still widely present. Thus, their virtual counterparts, being more subtle, pose a greater threat.

\section{Methodology} \label{sec:method}
In the following, we describe the experimental methodology for measuring the network of our university. We focus on monitoring the prevalence of mining in the institution.



Our university has several systems for detecting invasions and threats that are used to provide security for its students, faculty and staff. In this work, we consider data provided by two tools used by the University Information Security: SGIS -- System Incident Management System Security, provided by the National Education and Research Network and the management system of tickets used by the institution. The ticket system includes tickets requested by members of our university, while SGIS generates tickets using the result of network traffic scans at the PoP (point of presence at the state of Rio de Janeiro, encompassing many academic institutions).


Discovery of mining attacks begins with the notification of any irregularities. The notification of irregularities occurs in three ways, carried out in parallel: $(a)$ notification from partner institutions and government agencies informing that the hosts are communicating with other infected mining machines;
$(b)$ monitoring the university's network traffic by checking if the host is communicating with mining servers, also known as mining pools;
$(c)$ notifications from a machine   indicating anomalies in behavior.

From the notification, a physical audit is performed on the suspect machines to detect the possible attack vector involved. In this audit, data is also collected on the configurations of the machines involved. 
The audit  consists of:

\textit{           \textbf{Physical inspection.}} Physically  reaching  the machine, and analyzing   the purpose of the machine and active connections using Wireshark or
tcpdump, in conjunction with netstat (e.g., many active UDP connections for
a personal computer from a faculty in the History department can indicate
a suspicion that the machine has been compromised)

   \textit{           \textbf{Analysis of vulnerabilities.}}  Analysis of the vulnerabilities of the machine, from an external view, e.g., Shodan and from  an internal view, e.g., using
Apache vulnerability tracker provided by Kali Linux if  
installed on the machine

     \textit{             \textbf{Search for additional evidences.} } Search in the machine's folders (in particular, Apache subdirectories) for a
(binary) file that is typically used for illicit purposes. In our case,
we focus on searching for cryptocurrency mining files. So we do
a search in the names of all files on the machine for   strings  such as ``crypto'', ``Monero'' and ``Bitcoin''. We also  search for such strings in the system's cron for scheduling mining processes, as well as
as in the bashrc history

\textit{         \textbf{Analysis of browsing history.}} Search the history of sites accessed by the browser for suspicious sites. For  
example, in the case of suspected cryptocurrency mining, it is checked whether there are
records of access to publicly known mining sites, eg, http://btc.com.


The attack vectors that were found at our university   were: (i) virtual epidemic vectors (worms); (ii) virtual  targeted vectors (attacks on the network, coming from an Internet connection); (iii) physical vectors (attacks on the network, resulting from physical access to computers). In the particular case of cryptocurrency mining, physical access to machines represents an important vector.


\pgfplotstableread{./bars.txt}{\resdelay}

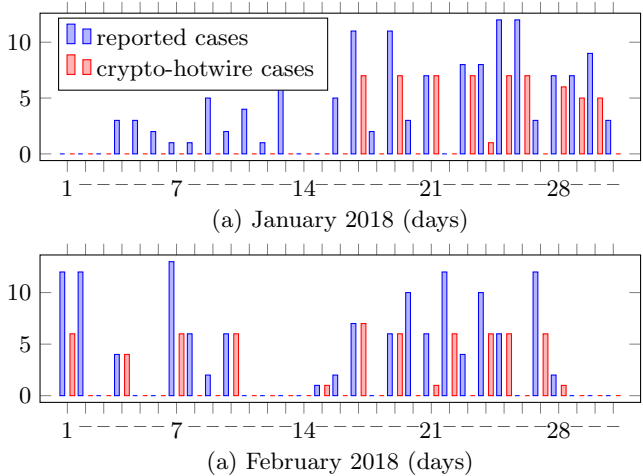
\begin{figure}[t!]
  \centering
  \begin{tikzpicture}
    \begin{axis}[width=0.54\textwidth, height=0.2\textwidth,
    ybar,
    bar width=1.8pt,        enlargelimits=0.05,
    xlabel={(a) January 2018 (days)},
    xtick=data,
    xticklabels from table={\resdelay}{A}, 
    legend pos=north west,
             legend cell align={left},   ]
    \addplot table [x expr=\coordindex, y=B]{\resdelay};
    \addplot table [x expr=\coordindex, y=D]{\resdelay};
    \legend{reported cases, crypto-hotwire cases} 
\end{axis}
  \end{tikzpicture}
   \begin{tikzpicture}
    \begin{axis}[width=0.54\textwidth, height=0.2\textwidth,
    ybar,
    bar width=1.8pt,        enlargelimits=0.05,
    xlabel={(a) February 2018 (days)},
    xtick=data,
    xticklabels from table={\resdelay}{A}, 
    legend pos=north west,
             legend cell align={left},   ]
    \addplot table [x expr=\coordindex, y=C]{\resdelay};
    \addplot table [x expr=\coordindex, y=E]{\resdelay};
\end{axis}
  \end{tikzpicture}
  \caption{Number of incidents (blue)   and crypto-hotwire incidents (red) between January 1 and February 28, 2018. } \label{fig:incidents}
  \end{figure}

\section{Prevalence of illegal mining} \label{sec:finding1}
The observed crypto-hotwire mining incidents occurred during the period from January 17 to February 28, 2018. In February 2018   mitigation strategies were deployed, and since then it has been more difficult to identify mining attacks.

In this study, we focus on general notifications received between January 1 and February 28, 2018. Figure~\ref{fig:incidents} reports the number of notifications of security incidents during the analyzed period. They cover notifications of various types of attacks, including DDoS, copyright infringement and brute force attacks on ssh. In Figure~\ref{fig:incidents} we notice that there are contiguous periods of greater  notification activity, for example, between January 25 and February 2, and between February 19 and February 25. In other contiguous periods, for example, between   January 1 and 8, and between  February  11 and 16, we have few notifications. The main reason for having these contiguous periods is related to the fact that when an attack occurs, the attacker   often manages to compromise multiple machines. When the problem is mitigated, the system is renewed and  switches to a new cycle.

Out of the reported incidents,  144 cases were classified as crypto-hotwire incidents. The dates of the notifications for these cases are reported in red bars in Figure~\ref{fig:incidents}. The notifications are daily, and the script that takes information from our university's backbone uses deep packet inspection and generates the notifications. The script runs throughout the day, and at midnight it restarts the  counter of notifications. This explains why, in almost all days, the number of notifications ranges from 7 to 6. Probably the devices compromised and the users running illegal mining were almost always the same, and their activity generated around 7 predictable notifications per day (e.g., in communication between the user and the mining center). It should be noted that on some days, for example, January 27 (Saturday) we have no notifications because our university network is unstable on weekends, undergoing maintenance in some of these periods. On certain days, for example, on January 24, we had a fall in the number of crypto-hotwire  notifications, which subsequently increased on 25 January. This is due to the the fact that on January 24 the mining problem was solved provisionally, reinstalling some machines in the system. However, the problem  returned as the attack vector was still active and resumed its activities on  January 25. Below, we describe in more detail one of the episodes that occurred during this period.

\textbf{Bitcoin mining episode.} In a unit of our university, we identified computers mining Bitcoin. From a  browser's history, we identified that a user had accessed mining pool sites, \texttt{http://btc.com} and \texttt{http://f2pool.com}. On these sites, the user logs in and leaves the computer on to raise funds from mining. Carelessly, the  user under consideration left a file with a trace of his logging and mining passwords on the machine. With the credentials of access to the mining pools, it was possible to identify the start and end times of each mining activity. It should be noted that the notification of the above event, which alerted the security to take  initiatives, was generated by the fact that there was frequent access to mining pool websites. Thus, suspicious traffic was generated. If mining had been done  executing  minimum  interactions with the rest of the network, we believe that it would have been much more difficult to detect mining activity.  

Among the most mined virtual currencies at our university, we highlight  Bitcoin and Monero. Bitcoin was the first virtual currency to be adopted on a large scale, which justifies its prevalence. Bitcoin is a reference for other cryptocurrencies, and there are already numerous simple programs to mine Bitcoin on any platform (Linux, MacOS, Android and Windows, in the most diverse versions). Monero~\cite{monero}   provides its users with privacy levels greater than those offered by Bitcoin. In addition, like Bitcoin, it also provides mining software facilities for script kiddies. For these reasons, many miners have adopted Monero as the first choice cryptocurrency, as identified in our measurements.

\section{Estimated gains and costs} \label{sec:finding2}
Next, we report results obtained with controlled experiments in the lab to estimate   gains and costs associated with mining at our university. In particular, we sought to emulate the   machines that we found to be mining in January and February 2018, as described in the previous section.

\textbf{Controlled experiment environment.} We consider three configurations that correspond to the machines that we have found to be mining Monero and Bitcoin    at our university. It is not surprising that the configurations are standard on computers widely used through our university, considering that equipment purchases are frequently made in batches. Thus, the configurations also capture  potential mining gains and costs across our university as a whole. 

The experiment counted with 3 machines. Configurations 1, 2 and 3 correspond to an Intel Pentium Dual Core, Intel Core i3 and Intel Core i5, all with 4 Gb of RAM and 2 cores except i5 which counts with 4 cores.   The hosts run the Linux Ubuntu 16.04 operating system, and the  energy consumption monitoring software PowerTOP. The mining software chosen was MinerGate, for the Monero virtual currency. For network data collection purposes (monitoring network flows) we use NFDump (netflow). 

\textbf{Experimental results. }
%
Let $r(j)$ be the hashing power of a machine with configuration $j$. In our measurements, we found $r(1) = 85 H/s, r(2) = 88 H/s,  r(3) = 177 H/s$, respectively.
In a system where $N$ machines have configuration $j$, the total hashing power of the system is given by $R(j)=N r(j)$, $j = 1,2,3$. In what follows, we consider $N=3$ which corresponds to approximately the number of machines found mining at our university  between January and February 2018.

The daily gain corresponding to the hashing power reported above, with the $N=3$ hosts, is given by $G(1)$ = R\$ 0.27, $G(2)$ = R\$ 0.28 and
$G(3)$= R\$ 5.61, under  configurations 1, 2 and 3 respectively. To obtain such values, we use the average quotation of   Monero in the Brazilian currency (reais) during the study period. It is worth noting that the expected gains are very small compared to costs, which are an order of magnitude higher, as indicated below. 

To estimate the energy expenditure per host, we measure the energy consumption of machines. The   measured energy consumption of the machines was 
$E^{(1)}=21$~W/h, 
$E^{(2)}=40$~W/h and
$E^{(3)}=40$~W/h,
for configurations 1, 2 and 3, respectively. Multiplying by the number of active hosts assumed to be 3, and for the cost of energy $c$ we obtain the estimated total cost, $
C^{(j)} = E^{(j)} N c$, with $C(j)$ equal to R\$ 61.69, R\$ 117.50 and R\$ 117.50 for configurations 1, 2 and 3, respectively.

\begin{figure}[t]
    \centering
    \includegraphics[width=0.5\textwidth]{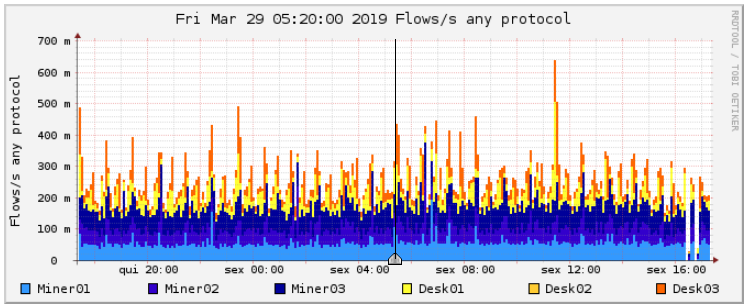}
    \caption{Traffic generated by miners is similar or smaller than standard desktops.}
    \label{fig:traffic}
\end{figure}

Until then, we have not taken into account the network patterns associated with mining in our controlled experiment, nor the underlying network costs. For a preliminary analysis  on network aspects,  we consider 3 additional machines  with the same configurations as the ones considered so far. Such additional machines are baseline machines, which run general-purpose desktop applications, and which do not mine. One of our purposes is to contrast the network usage of the corresponding machines.  As shown in Figure~\ref{fig:traffic}, the traffic pattern of hosts that are mining is lower than that of hosts performing standard equipment activities.  This fact corroborates the idea that it is more challenging to identify mining attacks remotely, as considered in this work, in comparison with local detection in the browser, using mine blockers tools~\cite{minerblock}.

\textbf{Monetization of mining for public benefit. }
The monetization of mining, to the benefit of the university itself, is potentially promising, \emph{but involves
ethical aspects and   environmental implications.}
 Since the university has periods of low movement or idleness of resources,  which causes resources to be underutilized, mining may be considered during these periods.  Predicting the value of cryptocurrencies and the surplus energy with which each institution can count a priori, however, is non-trivial. In domestic households in numerous cities in the Brazil, for example, there is a minimum energy consumption associated with the basic  minimum value of the electricity bill. In homes where this limit is not reached, or in underutilized commercial establishments, e.g., affected by the COVID-19 pandemic, it is possible to monetize the monthly energy surplus. 

\section{Conclusion} \label{sec:conclusion}
In this paper we have reported mining attacks   leveraging direct physical access   to computational resources.  Such   attack vectors are easy to implement in public infrastructures, and those attacks potentially generate financial resources that can be converted into any type of goods. We  presented measurements indicating that those mining attacks occurred  at our university. In particular, we indicate that Bitcoin and Monero were the two most mined coins.  We believe that this work opens up many directions for future developments. In particular, we envision measurements   to assess the prevalence of crypto-hotwire attacks beyond the university environment, e.g., in regions where energy theft is prevalent~\cite{BBC},    counting with collaborations between energy and communication service providers.  As another direction for future research, we envision new services leveraging the monetization of mining for public benefit, and leave their  ethical and environmental implications as subject for future work.

\bibliographystyle{acm}
\begin{small}

\bibliography{sample-base}

\end{small}

\end{document}